# Micro-electromechanical photonic integrated memristors

Matthew Zimmermann,[1] Julia M. Boyle,[1] Hardit Singh,[1] Alex Witte,[1] Kevin Palm,[1] Thuy-Linh Le,[1] Andrew J. Leenheer,[2] Daniel Dominguez,[2] Matt Eichenfield,[3,4,5] and Mark Dong[1,6]

*[1]The MITRE Corporation, 202 Burlington Road, Bedford, Massachusetts 01730, USA*
*[2]Sandia National Laboratories, P.O. Box 5800 Albuquerque, New Mexico, 87185, USA*
*[3]College of Optical Sciences, University of Arizona, Tucson, Arizona 85719, USA*
*[4]Electrical, Computer, and Energy Engineering, University of Colorado Boulder, Boulder, Colorado 80309, USA*
*[5]matt.eichenfield@colorado.edu*
*[6]mdong@mitre.org*

**Abstract:** Programmable optical memristors embedded in photonic integrated circuits (PICs) are emerging as an important technology for high-speed optical storage and in-memory optical computing applications. These devices provide multi-level, non-volatile storage of optical phases that can be interrogated at the speed of light, enabling parallel data readout or energy-efficient multiply-accumulate operations in artificial neural networks. However, there remains several outstanding challenges with existing optical memristor technology including durability, material-induced optical losses, large-scale reconfigurability, or fabrication yield for realistic applications. Here we introduce an analog-programmable photonic memristor based on photonic integrated micro-electromechanical (MEMS) cantilevers produced in a CMOS foundry. The memristor consists of low-loss silicon nitride waveguides, requires no additional back-end materials integration, and is all electrically programmed with electrostatic-piezoelectric forces. We demonstrate up to 5-bit phase storage levels, 50 kbit/s programming speeds, strain-assisted non-volatility lifetimes >1 hour, >1 billion cycle endurance, and stress-tested millions of write-read cycles with pseudorandom bit sequences. We further extend the memory lifetime to several days with simple electronic refresh circuits in a portable battery-powered module, demonstrating a proof-of-concept optical random-access memory in static or dynamic configurations. Our MEMS-photonics technology represents an important step toward practical optical memristors.

## 1. Introduction.

In order for integrated optical memristors [1-3] to achieve utility outside the laboratory, we require devices that have i) long-lived memory lifetimes, ii) high-speed, simple programmability, iii) error-free multi-bit operation, iv) high durability and endurance for a large number of switching cycles, and v) scalability to large number of devices with reliable fabrication. Many memristor technologies have met one or more of these requirements with great successes, although how best to achieve all desired metrics simultaneously is still an open question. For example, research on phase-change materials (PCMs) made of chalcogenides [4] shows highly promising results such as truly non-volatile "set-and-forget" operation with large refractive index contrasts. PCMs are programmable optically [5,6] and electronically [7,8] with reported applications to neural networks [9]. Outstanding challenges remaining include high optical losses, low device endurance for both electrical and optical programming approaches, and thermal engineering complexities [10] related to rapid material heating and cooling due to the high temperatures involved. Ferroelectric materials are another promising avenue for memristors. The major advantages include lower material optical losses, low electrical switching energies, and longevity of the induced ferroelectric domains reported in materials such as lead zirconate titanate (PZT) [11] and barium titanate (BTO) [12-14]. The quality of the films is also high enough yield for multiple devices to function on the same chip [11]. Some challenges are a typical DC bias requirement for optical readout in addition to hysteresis effects, which require magnetic repoling at high voltages. Other materials such as vanadium dioxide ($VO_2$) [15,16] and magneto-optics [17,18] have exhibited promising results, although high optical propagation losses are still present. Alternatively, integrated micro-electromechanical systems (MEMS) for optical memories have shown greater durability and scalability [19,20] with recent reports utilizing van der Waals forces for non-volatility [21]. MEMS modulation does not in principle require lossy or exotic materials to operate and can be implemented using electrostatic actuation fully compatible with existing CMOS fabrication processes. However, MEMS non-volatile memories are currently limited to binary switches [22-24] with existing devices requiring complex structures with challenging fabrication. Unlocking multi-bit, non-volatile phase storage using MEMS while retaining its inherent reliability advantages would be highly impactful in advancing optical storage and optical computing technologies.

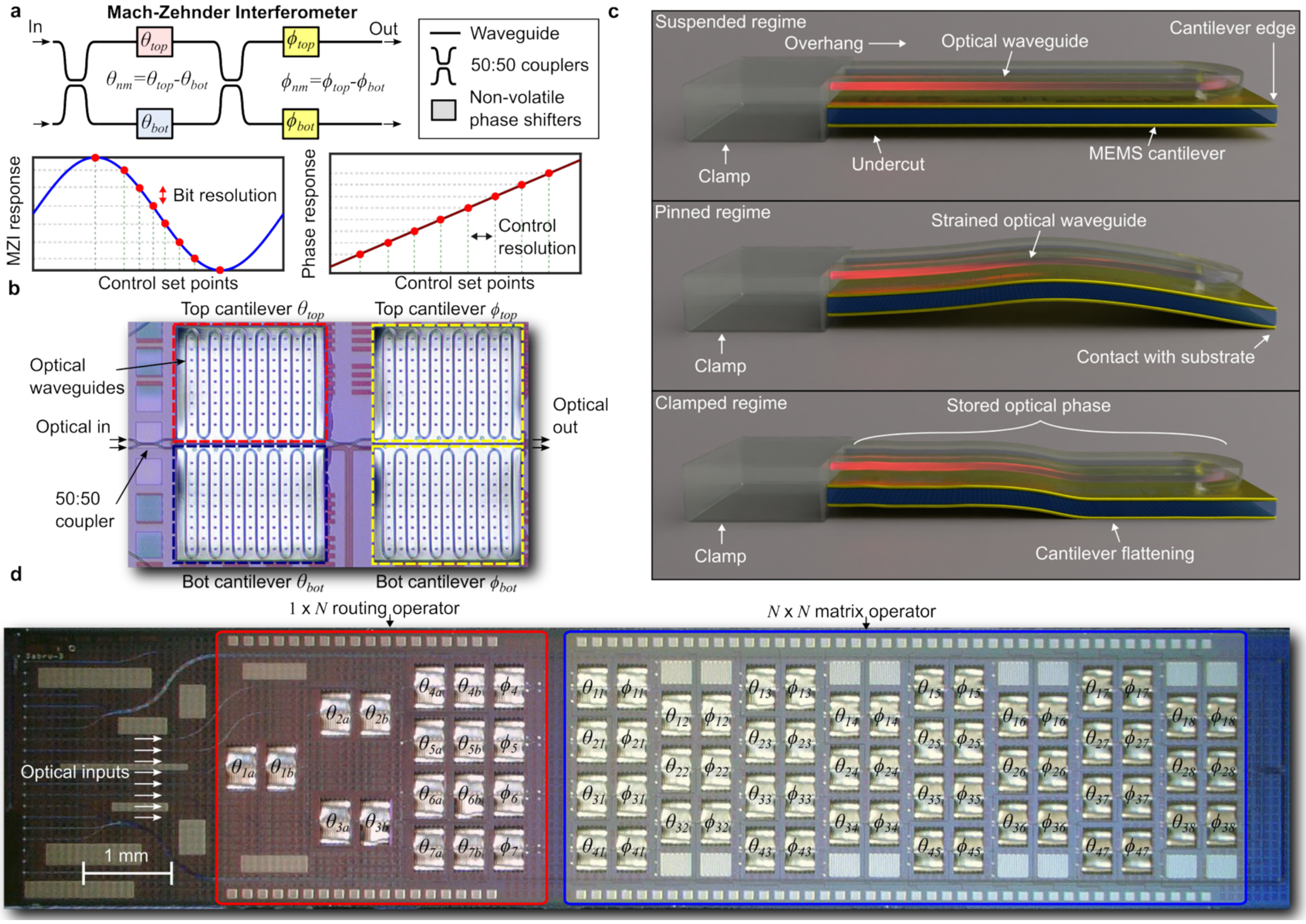

**Fig. 1: Micro-electromechanical photonic integrated memristors.** a) Diagram of a Mach-Zehnder interferometer (MZI) with long-lived memory phase shifters on each arm and output. The phase shifters can be programmed to execute a 2 x 2 unitary operation by tailoring the amplitude and phase response of the MZI to the programmable bit resolution. b) Optical microscope photo of an example MZI consisting of MEMS-cantilever phase shifters with labeled waveguides and 50:50 couplers. c) 3D conceptual render of each cantilever's mechanical phase storage mechanism, enabling multi-level memory based on the degree of substrate contact and deformation. d) Optical microscope photo of an example Mach-Zehnder mesh consisting of cascaded MZIs for optical routing for memory readout or matrix-vector multiplications implementing an in-memory compute architecture.

Here we introduce a MEMS-photonic-integrated memristor capable of accurate, durable, and multi-bit phase storage with simple electrical programming protocols, all fabricated on a fully CMOS-compatible layer stack [25]. Fig. 1 outlines the core concept of our photonic memristor. Fig. 1a shows the basic operation of the memristor in phase and amplitude configurations implemented by a Mach-Zehnder interferometer (MZI). Each phase shifter embeds a programmed phase in the MZI structure which produces a 2 x 2 unitary transfer function [26] when read out with an optical signal. The core geometry of each phase shifter is a singly clamped cantilever with meandering waveguides whose modulation stems from piezoelectric and electrostatically induced strain. Fig. 1b shows an optical microscope image of an exemplary MZI using cantilevers. Multi-level phase storage is induced by pulling the cantilever and optical waveguides to the desired state (3D render shown in Fig. 1c). This phase is retained by a mixture of electrostatic charge retention and nonlinear mechanical effects such as buckling and partial stiction to the substrate. With such a device on hand, Fig. 1d illustrates conceptually the large-scale photonic memristor system where a cascade of cantilever modulators stores phases for in-memory optical computing or optical storage applications. The 1 x *N* routing circuit would enable a solid-state optical storage medium while the *N* x *N* mesh imparts a stored matrix-vector multiplication implementing an in-memory compute architecture [27] applicable for artificial neural networks.

Through endurance test experiments, we show that through voltage control pulses, the cantilever can be accurately and repeatably programmed into several long-lived memory states. The memory decays slowly with a strain-assisted natural lifetime of 4000 seconds or more but is easily refreshed by a reapplication of the same control pulse or simply reprogrammed to a different state. Multi-bit operation is then further stress-tested for tens of thousands of read-write cycles under pseudorandom bit sequences. We lastly show the practicality of the memristor by extending the phase storage with a battery-powered, packaged prototype with simple electronic refresh circuits built from commercial off-the-shelf components. The details of the device design and performance are in the following sections.

## 2. Device Design and Programming

We show the photonic memristor device design and basic performance in Fig. 2. The primary test device is an MZI with two 50:50 directional couplers and two cantilever phase shifters, one on each arm (Fig. 2a). The optical layer consists of low-loss silicon nitride (SiN) waveguides (-0.4 dB/cm [28]), cladded with silicon oxide. Fig. 2b displays a scanning electron microscope (SEM) image of the cross section of the cantilever. MEMS actuation of the cantilevers is achieved by strain and geometric deformation provided by electrostatic-piezoelectric forces [29]. The electrical signals are implemented using three distinct metal layers, M1, M2, and M3, with M1 and M2 forming the electrostatic control and M2 and M3 forming the piezoelectric control. The electrostatic force strength is facilitated by the thickness of the oxide and amorphous silicon (a-Si) layers while the piezoelectric forces are generated by applying a field to the aluminum nitride (AlN) material. These two independent voltage controls enable both piezo- and electrostatic-induced deformations of the cantilever structure. These deformations propagate to the optical waveguiding layers, thereby inducing effective refractive index shifts and programming the MZI to the desired output state.

For basic experimental characterization, we coupled continuous-wave light from a 730 nm laser using a fiber array and on-chip grating couplers while applying different voltages to the MEMS controls using an RF probe. We measured the cross-port transmission of the MZI under DC electrostatic and DC piezoelectric actuation (Fig. 2c and 2d, respectively) to validate the modulation of each single-arm cantilever. The piezo response is linear with voltage, showing a ~33 dB extinction while the electrostatic is nonlinear with voltage, driving strongest in the 30 – 50V regime and reaches ~25 dB extinction, both in line with previously reported cantilever devices [29,30]. To confirm the actual deformation, we performed white-light profilometry of the cantilever under electrostatic DC biases of 0 V and 40 V, as demonstrated in Fig 2e. We observed in the rest position (0 V), the cantilever was curled up and suspended from the bottom substrate. As we increased the DC bias to 40 V, the cantilever contacted and flattened significantly along the bottom oxide substrate. This flattening allows the cantilever to stick to the bottom substrate even after the DC bias was removed, allowing the mechanical state, and hence the phase, to be retained. The mechanical retention was then verified by switching the cantilever into a high-impedance (high-Z) state while biased at 40 V. The deformation of the cantilever was observed with a white-light profilometer immediately, one minute, and two minutes after being switched into high-Z. Fig. 2f shows a representative 1D cut across the cantilever at 0 V, 40 V, and in the high-Z state at the aforementioned times. Even after two minutes of the cantilever being in the high-Z state, the cantilever still retained the 40 V state shape, confirming the slow mechanical relaxation time.

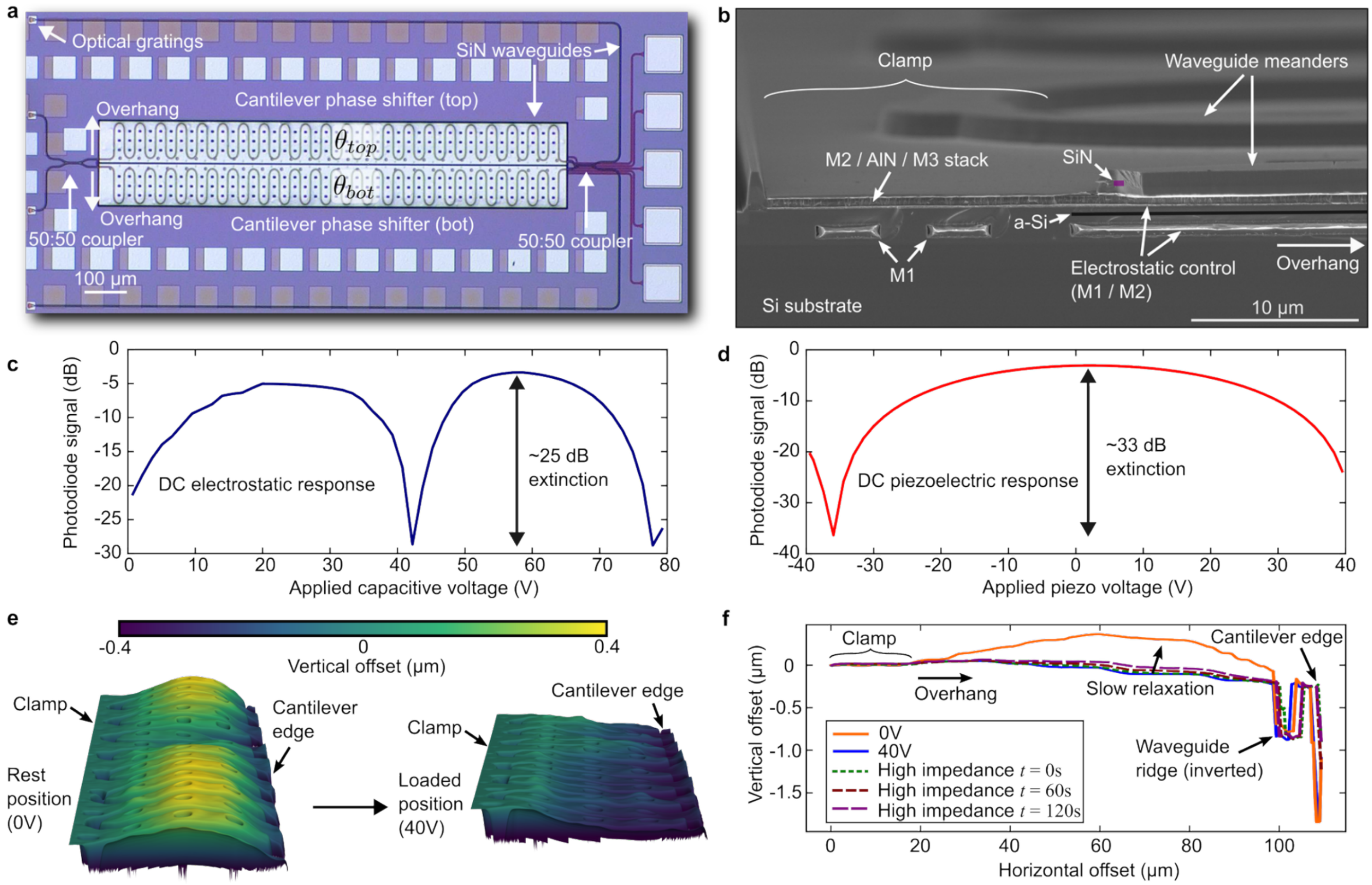

**Fig. 2: Design and basic characterization of a MEMS-cantilever photonic memristor.** a) Optical microscope image of the MZI test structure with waveguides, 50:50 couplers, and cantilever phase shifters labeled. b) SEM cross section showing the PIC materials layer stack (SiN and a-Si are false colored purple and black, respectively). Waveguide meanders on the overhang section relative to the cantilever's clamp are labeled. c) DC electrostatic and d) DC piezoelectric response curves showing 25 dB – 33 dB optical power extinction. e) White-light profilometer measurements of the cantilever's surface deformation at rest and with 40 V applied to the capacitive electrodes, showing a flattening behavior. f) 1D line plots of the profilometer data at 0 V, 40 V, and high impedance (disconnecting) after 40 V, showing a slow relaxation back to the original shape.

From these basic observations, the operation of our photonic memristor is as follows. Writing data involves applying a voltage to the electrostatic electrodes of one of the cantilever arms of the MZI, waiting for the deformation to complete, and disconnecting to high-Z, thus locking in an optical phase state in the mechanical strain. Reading data involves measuring the transmission through the MZI, determining normalized amplitude or unwrapped phase values, and decoding them back to binary. We outline the device's programming and optimal operation in Fig. 3, illustrating the shaded (high-Z) portions of operation. We first determined the decay lifetime by measuring the MZI's normalized transmission (Fig. 3a) after applying a drive voltage to electrostatic electrodes and disconnecting to a high-Z state in which power is no longer consumed. We then measured a slow transmission decay back to the original state (Fig. 3b). To mitigate the effects of the MZI's periodic transfer function, we repeated the experiment and plotted the unwrapped phase $\theta$ (Fig. 3c) whose decay was fitted to an exponential with natural lifetime $\tau = 4000$ s (Fig. 3d). While the phase offers a viable path for encoding and decoding information, the unwrapping process is less straightforward than direct detection of the optical field. For the remaining experiments in this paper, we opted for amplitude encoding via normalized optical intensity measurements and biasing the MZI's operating point with the phase shifter on the unused arm. Due to the mechanical separation of these cantilevers, we did not observe any crosstalk between the MZI arms [25].

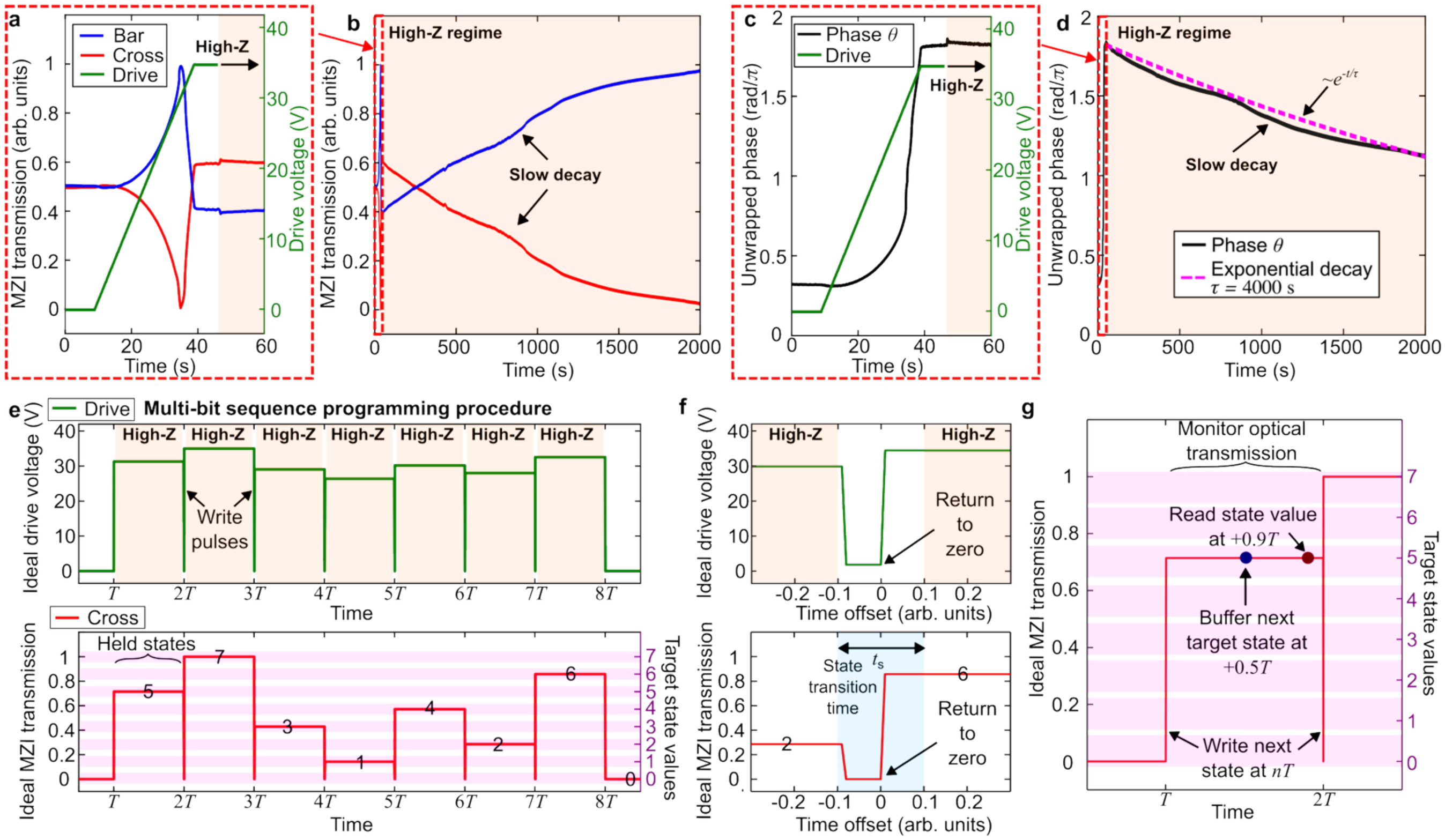

**Fig. 3: Photonic memristor operation and programming.** a) Normalized MZI transmission of the bar and cross states when a drive voltage is applied and then disconnected to high-Z (shaded region). The output levels maintain their states after the control voltage is disconnected. b) Normalized MZI transmission of the bar and cross states over a long duration in the high-Z state. The output levels slowly decay back to the original, unloaded state. c) Unwrapped phase from the MZI transfer function processed from the same data showing the effects of the drive voltage and high-Z regimes. d) Long-duration phase decay with overlaid decaying exponential with lifetime $\tau = 4000\ s$. e) Example of an ideal drive voltage (top) and idealized MZI transmission output (bottom) of a sequence of pulses in a 3-bit system, illustrating the procedure to program and read a sequence states. The high-Z regimes and target state value amplitude ranges are highlighted. We note that here during high-Z, the drive voltage is drawn as its most recent "set" value. f) Zoomed time trace of the example programming procedure with $t_s$ labeled. The return-to-zero (RZ) protocol is shown where the memristor is set briefly to 0 V for ~100 ms to improve memory fidelity. g) The timing of the buffer and read-state during a single read-write period.

Fig. 3e-g illustrate an exemplary programming protocol for 8-state (3-bit) operation using amplitude encoding. The timing of the programming protocol is determined by the total read-write repetition period $T$ and the cantilever's lifetime $\tau$ (Fig. 3e). Intuitively, the natural lifetime approximately determines the longest possible time between either a refresh pulse or a programming to a different state (which will depend on the number of operating bits for highest fidelity readout). Conversely, the state transition time $t_s$, defined as the total time it takes for the memristor to switch from one state to another state (Fig. 3f), will determine the fastest programming rate. For our experiments, the choice of $t_s$ is effectively narrowed to two options: 1) We simply program the cantilever as quickly as possible by switching the voltage directly to the new setting for the next state. This is eventually limited by the cantilever's RC constant, on the order of ~10 µs. 2) For slower programming applications, we implement a "return-to-zero" (RZ) protocol where the control voltage is reset back to 0 for ~100 ms before

switching to the next state such that $t_s \sim 200$ ms. Because of a small mechanical hysteresis in the cantilever, the RZ protocol greatly improves memory fidelity when operating at higher bit memory levels. Lastly, to evaluate the fidelity of the read-write process, we monitor the optical transmission for the entire bit sequence. Fig. 3g shows the timing of the write-read operations. The computer buffers the next target state at half the read-write repetition period to prepare the voltage controller (arbitrary waveform generator) for the write process. The memristor is “read” by sampling the optical power after roughly 90% of the read-write repetition period has elapsed to determine which state was actually programmed. The sampled value is then decoded and checked against the desired target state, logging any bit errors should they occur. Once we determined the full programming protocol, the important metrics of memristor throughput and energy consumption can then be evaluated.

### 3. Binary and Multi-Bit Endurance Experiments

To implement the programming protocol, our experimental setup is as follows. For low-rate binary and multi-bit endurance tests, we connected the electrostatic electrodes to an electromechanical single-pole, double-throw (SPDT) RF switch that would mechanically switch between the drive signal (0 V - 35 V) and a high-impedance open circuit. We used a gate pulse of 200 ms and a gate voltage of 28 V for the SPDT switch. Both the drive signal and gate pulse were generated by a multi-channel arbitrary waveform generator (AWG) and amplified by a high-voltage amplifier with 20x gain. We lastly used an additional amplified voltage channel to set the opposite MZI arm’s cantilever (using the piezo modulator) to a pre-calibrated, static DC voltage. This biased the MZI transfer function to be in a maximum cross state transmission when the capacitive portion is set to 0 V, enabling full use of the modulation range. For executing the RZ protocol in Fig. 3f, we first set the drive signal to 0 V while the SPDT switch was still in high-Z before the state transition occurred. Once the 200 ms gate was applied to the SPDT switch, the cantilever was “cleared” to 0 V for the first 100 ms, then the drive signal was set to the target value for the next state. After the second ~100 ms elapsed, the gate pulse returned to zero thus switching the cantilever back to high-Z. This process left the next state voltage loaded onto the cantilever. Finally, after another 100 ms after the switch to high-Z, the drive signal returned to 0 V to reduce unnecessary power draw from the driving signal’s AWG.

For high-rate drive pulse sequences (i.e. very short read-write duration times *T*) or certain binary experiments, the RZ protocol was not used due to the switching time of the SPDT switch being a limiting factor and the RZ protocol being unnecessary for high-speed pseudorandom binary operation. For sequences with >10 kHz pulse repetition rate, we held the switch closed for the duration of the experiment while the dynamic voltage control occurred through the AWG only. Throughout the entire programming procedure, the memristor’s optical outputs were actively captured on high-speed photodiodes and logged over the duration of each experiment.

We stress-tested our photonic memristor with a series of binary and multi-bit endurance and accuracy experiments. For these tests, we generated a pseudorandom state sequence and loaded them into the AWG. The states were converted to drive pulse values that ran synchronized with the gate pulses (see Supplementary Section 1 for calibration values). For each state, we sampled the bar and cross transmissions prior to the next transition and calculated the normalized power $P_{bar,cross}^{norm} = P_{bar,cross}/(P_{bar} + P_{cross})$. Finally, we decoded the sampled power $P_{sampled}^{norm}$ (using either cross or bar states) to retrieve a state or bit value and compared the result to the transmitted pseudorandom state sequence. To calculate the bit error rate, we assigned each state to a Gray code sequence and compared our transmitted sequence to our received sequence.

We plot the results of our binary endurance experiments in Fig. 4. The binary states 0 and 1 are defined as simply the top half ($P_{sampled}^{norm} > 0.5$) and bottom half ($P_{sampled}^{norm} < 0.5$) of the normalized intensity. We plot the measured time trace of the cross state ($P_{cross}^{norm}$) for a long read-write pulse period $T = 1000$ s in Fig. 4a, where the state decay is clearly seen. Here we programmed 50 states in a strictly alternating binary sequence with sampled powers values shown in Fig. 4b. We observe some variability in the decay rate, with some samples close to the 0.5 threshold but overall, the bit error rate was 0 (Fig. 4c). We repeated the test with a much shorter read-write pulse period $T = 10$ s in Fig. 4d and with a full pseudorandom state sequence. Here the sampled power is highly accurate (Fig. 4e) due to minimal phase decay, again logging zero errors for over 8000 bits transmitted (Fig. 4f). We again repeated the test with an even shorter read-write pulse period $T = 40$ µs (25 kHz repetition rate) in Fig. 4g. Without the RZ protocol, the memristor switches as quickly as its capacitor can charge. The sampled states in Fig. 4e show good accuracy well within the state 1 region. We logged no errors in over 117 million bits transmitted (Fig. 4i), showing great robustness and endurance in binary operation. Lastly, we ran a high-speed $T = 20$ μs (50 kHz repetition rate) endurance test for over 1 billion cycles. We plot the memristor's switching performance after the billion-cycle test in Fig. 4j showing no noticeable decay or drop in operating speed and accuracy.

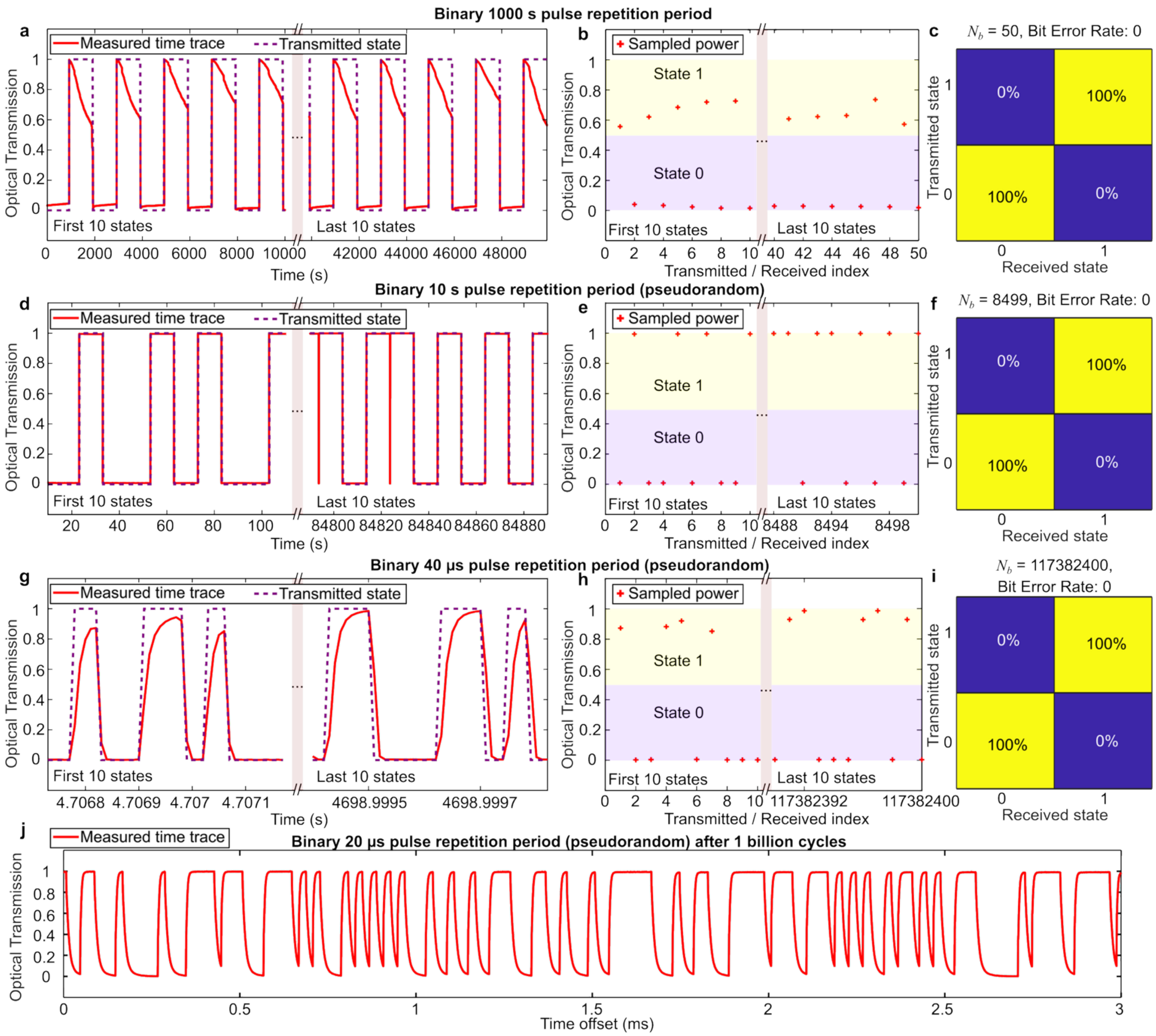

**Fig. 4: Binary endurance and fidelity experiments.** a)-c) The measured time trace, sampled optical power, and transmitted / received error matrix respectively of an alternating binary sequence at 1000 s read-write period for a total of 50 bits transmitted at 0 bit errors. For this particular memristor, this represents close to the upper limit of re-programming time or refresh pulse spacing. d)-f) The measured time trace, sampled optical power, and transmitted / received error matrix respectively of a pseudorandom binary sequence at 10 s read-write period for a total of 8,499 bits transmitted at 0 bit errors. At more frequent repetition rates, the sampled power values are much closer to their ideal values due to less time for phase decay. g)-i) The measured time trace, sampled optical power, and transmitted / received error matrix respectively of a pseudorandom binary sequence at 40 μs read-write period (without the RZ protocol) for a total of 117,382,400 bits transmitted at 0 bit errors. j) Measured time trace of the optical memristor of a pseudorandom binary sequence at 20 μs read-write period (without RZ protocol) after a >1 billion binary cycle endurance test. The time trace shows no observable degradation and still operates at 0 bit errors. This represents the lower limit of programming or refresh pulse spacing and corresponds to a 50 kbit/s maximum programming rate.

Moving to multi-bit operation, we show our experimental results in Fig. 5. Starting with four states (2 bits), we applied a pseudorandom state sequence at a 10 s read-write period (Fig. 5a) using the same sampling timing discussed in Fig. 3

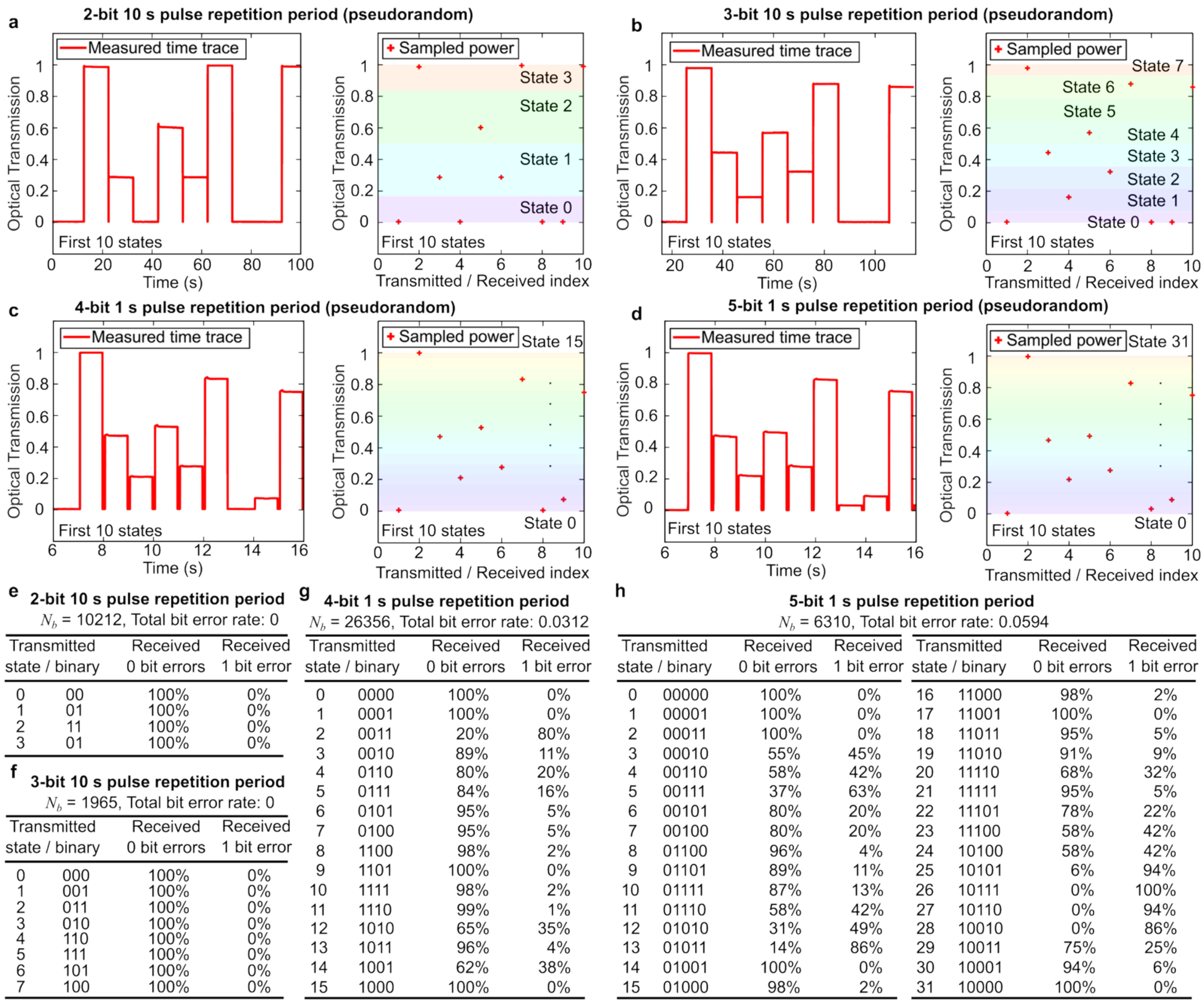

**e** 2-bit 10 s pulse repetition period

$N_b$ = 10212, Total bit error rate: 0

| Transmitted state / binary | | Received 0 bit errors | Received 1 bit error |
|---|---|---|---|
| 0 | 00 | 100% | 0% |
| 1 | 01 | 100% | 0% |
| 2 | 11 | 100% | 0% |
| 3 | 01 | 100% | 0% |

**f** 3-bit 10 s pulse repetition period

$N_b$ = 1965, Total bit error rate: 0

| Transmitted state / binary | | Received 0 bit errors | Received 1 bit error |
|---|---|---|---|
| 0 | 000 | 100% | 0% |
| 1 | 001 | 100% | 0% |
| 2 | 011 | 100% | 0% |
| 3 | 010 | 100% | 0% |
| 4 | 110 | 100% | 0% |
| 5 | 111 | 100% | 0% |
| 6 | 101 | 100% | 0% |
| 7 | 100 | 100% | 0% |

**g** 4-bit 1 s pulse repetition period

$N_b$ = 26356, Total bit error rate: 0.0312

| Transmitted state / binary | | Received 0 bit errors | Received 1 bit error |
|---|---|---|---|
| 0 | 0000 | 100% | 0% |
| 1 | 0001 | 100% | 0% |
| 2 | 0011 | 20% | 80% |
| 3 | 0010 | 89% | 11% |
| 4 | 0110 | 80% | 20% |
| 5 | 0111 | 84% | 16% |
| 6 | 0101 | 95% | 5% |
| 7 | 0100 | 95% | 5% |
| 8 | 1100 | 98% | 2% |
| 9 | 1101 | 100% | 0% |
| 10 | 1111 | 98% | 2% |
| 11 | 1110 | 99% | 1% |
| 12 | 1010 | 65% | 35% |
| 13 | 1011 | 96% | 4% |
| 14 | 1001 | 62% | 38% |
| 15 | 1000 | 100% | 0% |

**h** 5-bit 1 s pulse repetition period

$N_b$ = 6310, Total bit error rate: 0.0594

| Transmitted state / binary | | Received 0 bit errors | Received 1 bit error | Transmitted state / binary | | Received 0 bit errors | Received 1 bit error |
|---|---|---|---|---|---|---|---|
| 0 | 00000 | 100% | 0% | 16 | 11000 | 98% | 2% |
| 1 | 00001 | 100% | 0% | 17 | 11001 | 100% | 0% |
| 2 | 00011 | 100% | 0% | 18 | 11011 | 95% | 5% |
| 3 | 00010 | 55% | 45% | 19 | 11010 | 91% | 9% |
| 4 | 00110 | 58% | 42% | 20 | 11110 | 68% | 32% |
| 5 | 00111 | 37% | 63% | 21 | 11111 | 95% | 5% |
| 6 | 00101 | 80% | 20% | 22 | 11101 | 78% | 22% |
| 7 | 00100 | 80% | 20% | 23 | 11100 | 58% | 42% |
| 8 | 01100 | 96% | 4% | 24 | 10100 | 58% | 42% |
| 9 | 01101 | 89% | 11% | 25 | 10101 | 6% | 94% |
| 10 | 01111 | 87% | 13% | 26 | 10111 | 0% | 100% |
| 11 | 01110 | 58% | 42% | 27 | 10110 | 0% | 94% |
| 12 | 01010 | 31% | 49% | 28 | 10010 | 0% | 86% |
| 13 | 01011 | 14% | 86% | 29 | 10011 | 75% | 25% |
| 14 | 01001 | 100% | 0% | 30 | 10001 | 94% | 6% |
| 15 | 01000 | 98% | 2% | 31 | 10000 | 100% | 0% |

**Fig. 5: Pseudorandom state sequence experiments for multi-bit memristor operation.** a) Measured time trace and first 10 sampled optical powers of the photonic memristor operating in 2-bit mode (4 states) with 10 s pulse repetition rate. b) Measured time trace and first 10 sampled optical powers of the photonic memristor operating in 3-bit mode (8 states) with 10 s pulse repetition rate. c) Measured time trace and first 10 sampled optical powers of the photonic memristor operating in 4-bit mode (16 states) with 1 s pulse repetition rate. d) Measured time trace and first 10 sampled optical powers of the photonic memristor operating in 5-bit mode (32 states) with 1 s pulse repetition rate. e)-h) Transmitted / received error tables for 2-bit, 3-bit, 4-bit, and 5-bit experiments respectively using Gray code sequences. The photonic memristor has error-free operation up to 3 bits, a 3.12% BER at 4-bit operation at a maximum error distance of 1, and a 5.94% BER at 5-bit operation with the vast majority of bit errors at distance 1 and a maximum error distance of 2 (states 12, 27 and 28).

(including the RZ protocol). The same is repeated for eight states (3 bits) in Fig. 5b. We also ran 16 states (4 bits) in Fig. 5c, and finally 32 states (5 bits) in Fig. 5d. We defined the multi-bit state levels by maximizing the distance between states (for example, 2-bit operation defines states at 0, 0.33, 0.67, and 1.0 normalized transmission). The sampled power is decoded by simply rounding to the closest valid state (colored boxes illustrate the boundaries in the sampled power plots).

Several thousands of bits were transmitted for all tests whose error rates and values are shown in Fig. 5e-h (2, 3, 4 and 5 bits respectively). Generally, the memristor operated well at lower bit values but increased in bit error rate (BER) as the number of states increased. We observed no errors for 2-bit and 3-bit operation (after 10212 bits and 1965 pseudorandom bits transmitted, respectively). Using a gray code encoding, we measured a BER of ~3.1% in 4-bit operation for 26356 pseudorandom bits transmitted and a BER of ~5.9% in 5-bit operation for 6310 pseudorandom bits transmitted. Despite the presence of errors, Fig. 5g shows all 4-bit errors were within an error distance of 1 state, while Fig. 5h shows only three states having any errors with distance greater than 1, suggesting existing telecommunications techniques [31] may be used to improve performance such as optical performance monitoring, drive voltage adjustment, and error correction codes (Hamming codes). Moreover, the error tables suggest the errors are not random but confined to certain states that have a high propensity for error (examples include states 25-28 in 5-bit operation). This was primarily due to difficulty in calibrating the memristor to achieve those specific transmission values due to the nonlinear mechanical nature of the electrostatic actuation. Different state definitions such as increasing state densities around 0 and 1 transmission, where actuation is more accurate, may reduce the observed BER. For these faster repetition rate multi-bit tests, we note the timing of when to sample the memristor's state (when 90% of the read-write period has elapsed, Fig. 3g) does not significantly affect the error rates.

## 4. Battery-powered Photonic Module

While the memory lifetime may be sufficient for fast reconfiguration tasks such as optical computing, other applications may desire a longer phase retention time. Taking inspiration from electronic static and dynamic random-access memories, we demonstrated a portable, battery-powered optical memristor module (Fig. 6). Fig. 6a shows the fully packaged photonic integrated memristor. The PIC was first attached to a PCB enabling piezo and capacitive voltages delivery via wirebonds to on-chip pads. Next, a fiber array was aligned and glued the PIC, completing the optical I/O. We combined the photonic package with a circuit designed for supplying the capacitive and piezo voltages using series of 9 V batteries. We used simple voltage dividers, adjustable with trimpots, to generate the required settings for the 0 and 1 binary states. For the refresh circuitry, we used a 555 timer and an electromechanical relay pre-calibrated to give an approximately 7 second refresh time. The full circuit details are described in Supplementary Section 1.

Fig. 6b-d plot the results of the battery-powered experiment. We first disconnected the refresh circuitry and performed a simple battery-powered hold test of the memristors over an 8-hour span. The optical outputs held well for binary operation (Fig. 6b) but showed some non-ideal decay near the end due to slowly decreasing battery voltages. The transition is also due to the MEMS cantilever exhibiting sharp mechanical behavior in that phase regime. To mitigate this, we programmed the

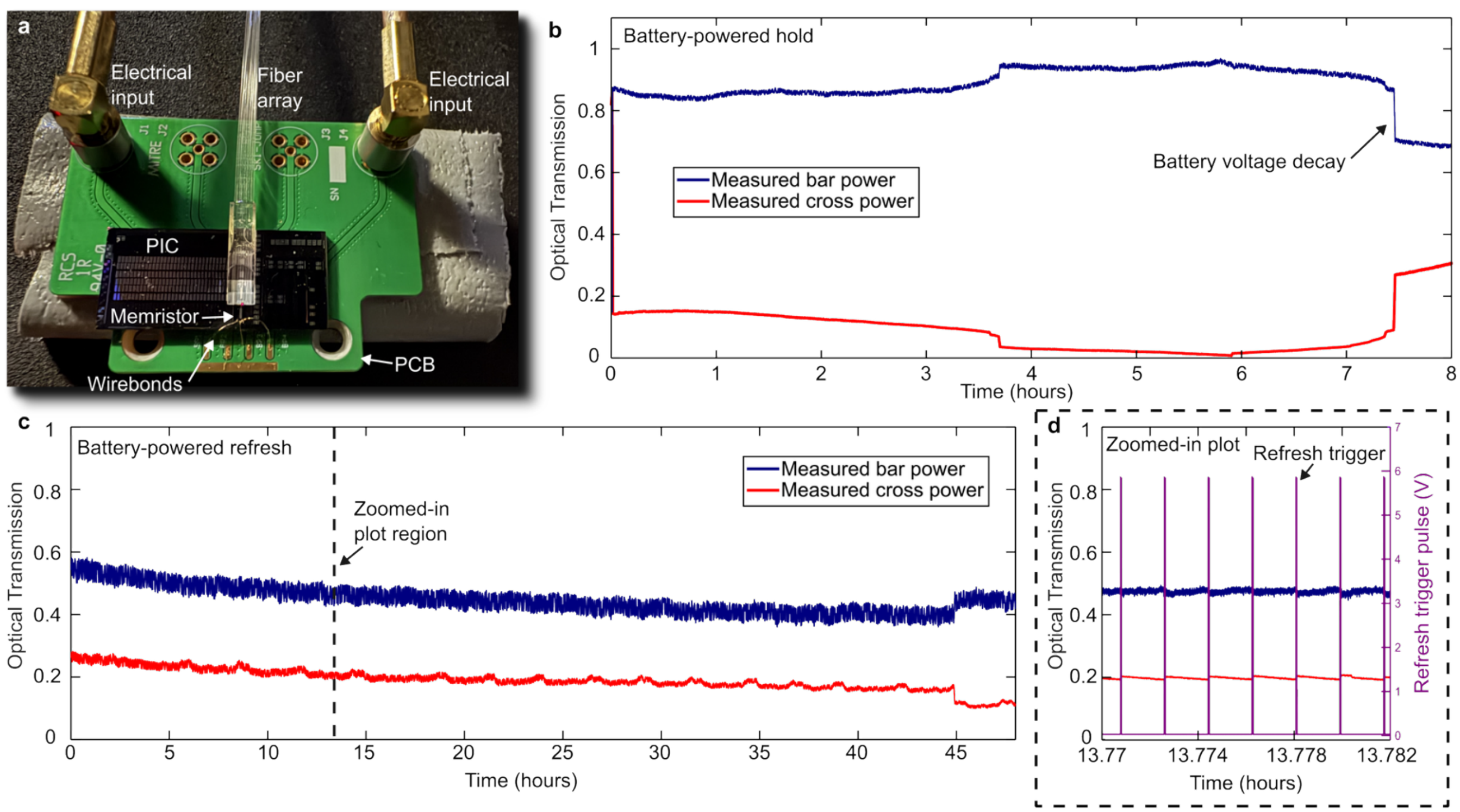

**Fig. 6: Battery-powered photonic integrated memristor.** a) Photo of the packaged photonic circuit with labeled electrical and optical I/O. b) Plot of the normalized bar and cross optical transmission with the batteries simply holding the piezo and capacitive voltages. c) Plot of the normalized bar and cross optical transmission with the refresh circuit active. d) Zoomed-in plot around hour 13 showing the measured refresh trigger pulse generated from the 555 timer overlaid with the optical transmission.

memristor to a more stable phase state for the battery-powered refresh experiment. Fig. 6c shows the monitored optical outputs over 48 hours remaining stable and distinct for binary decoding. Fig. 6d plots a zoomed-in region of Fig. 6c with the train of voltage pulses measured from the 555 timer. The pulse train triggers the electromechanical relay to reconnect the battery voltages to the photonic memristor, enabling a refresh operation. We add that the data shown in Fig. 6 were taken in open loop; there was no temperature control, trimpot adjustment, laser adjustments, or refresh circuit changes.

Conceptually, we may utilize the module by first connecting it to a computer to program specific optical phases in the MEMS devices. Then the module is disconnected, storing the phase values with strain potential and battery backup capabilities. Finally, after the desired storage time, the module is re-connected to a computer to optically read out the stored data.

## 5. Discussion

In summary, we report a simply programmable photonic memristor with great endurance and accurate multi-level storage fabricated directly with CMOS foundry tools. There are still several areas of improvement for our MEMS device. The first is the device footprint is quite large (~0.12 $mm^2$ ) and would benefit from fewer waveguide meanders and stress engineering [32] to improve mechanical deflection. This may also reduce actuation voltage and switching energy by virtue of a smaller device. Additional studies also are required to fully understand the nature of the stiction responsible for the long-lived memory lifetime, thus enabling future designs to improve the phase retention (the best of which we measured currently exceeds 20 hours). Based on current device measurements, we theorize a combination of mechanical potential [29] and van der Waal's forces [21] are responsible. Please see Supplementary Section 2 for additional experimental data elucidating device performance and behavior. The battery-powered circuit certainly increases the memory lifetime but could be easily improved with application-specific integrated circuits or CMOS voltage controllers. Additional areas of improvement are optical losses per modulator, which still hover around -1.5 dB per phase shifter with more recent fabrication runs improving to better than -1 dB. Here, the optical material (silicon nitride) is not the inherent limiting factor but instead the fabrication process which introduces hillocks and particulates above the cantilever, inducing scattering. Further refinements to the processing steps would alleviate these losses and approach SiN material losses.

Based on demonstrated past fabrication yield and reliability [30], it may be plausible to reach ~1 kbit optical storage in the near term (with main bottlenecks in electronic I/O and packaging). While the capacity is not comparable to conventional electronic memories, these optical memories may benefit certain applications where readout latency takes highest priority. Moreover, realistically scaling to thousands of stored bits would require co-integrated CMOS circuitry. Recent progress on this area has been promising [33].

Regarding energy use in computing applications, we comment a few observations. Due to the high endurance and reliability, high-speed inferencing with multiple trained models is possible but may not yield energy savings depending on the operation. The ideal regime for our memristor is a reconfiguration time of 10 – 100 seconds, whether by model switching and weight reprogramming, system recalibration, or simple refresh for continuous inference. Any shorter reconfiguration time will be dominated by programming energy which reduces the overall utility of zero-hold-energy memristors. Any longer reconfiguration time suffers from lack of inference model flexibility, system resilience to drift, and eventual voltage-refresh energy use becoming significant relative to the initial programming energy. We present a more detailed analysis comparing our technology to thermo-optic [34] and PCM modulators in Supplementary Section 3.

**Funding.** MITRE Quantum Moonshot Project; NSF FuSe 2 Award ECCS-2425611; U.S. Department of Energy, Office of Science; ARPA-E award DE-AR0002064.

**Acknowledgments.** M.D. thanks M. Saha for his technical assistance. M.D. also thanks D. Englund, G. Gilbert, and R. Han for their support. Sandia National Laboratories is a multimission laboratory managed and operated by National Technology & Engineering Solutions of Sandia, LLC, a wholly owned subsidiary of Honeywell International Inc., for the U.S. Department of Energy's National Nuclear Security Administration under contract DE-NA0003525. This paper describes objective technical results and analysis. Any subjective views or opinions that might be expressed in the paper do not necessarily represent the views of the U.S. Department of Energy or the United States Government.

**Author Contributions.** M.Z. and J.M.B. led the experimental work and performed the experiments. H.S., A.W., K.P., and T.L. provided further assistance in the experiments. M.Z. conceived the programming protocols. J.M.B. and M.D. designed the devices. H.S., M.Z., and M.D. performed the theoretical analysis. D.D. and A.J.L. implemented processing variations and supervised the device fabrication. M.D. conceived the device. M.D. and M.E. supervised the project. M.D. wrote the manuscript with input from all authors.

**Disclosures.** The authors declare no conflicts of interest.

**Data availability.** Data underlying the results presented in this paper are not publicly available at this time but may be obtained from the corresponding author upon reasonable request.

**Supplemental document.** See Supplement for supporting content.

# Micro-electromechanical photonic integrated memristors: Supplement

Matthew Zimmermann,[1] Julia M. Boyle,[1] Hardit Singh,[1] Alex Witte,[1] Kevin Palm,[1] Thuy-Linh Le,[1] Andrew J. Leenheer,[2] Daniel Dominguez,[2] Matt Eichenfield,[3,4,5] and Mark Dong[1,6]

*[1]The MITRE Corporation, 202 Burlington Road, Bedford, Massachusetts 01730, USA*
*[2]Sandia National Laboratories, P.O. Box 5800 Albuquerque, New Mexico, 87185, USA*
*[3]College of Optical Sciences, University of Arizona, Tucson, Arizona 85719, USA*
*[4]Electrical, Computer, and Energy Engineering, University of Colorado Boulder, Boulder, Colorado 80309, USA*
*[5]matt.eichenfield@colorado.edu*
*[6]mdong@mitre.org*

### S1. Experimental methods

**Experimental setup for memristor characterization tests:** Our experimental setup for PIC characterization is shown in Fig. S1a. The basic setup involves delivery and retrieval of optical signals via a 4-port fiber array aligned to four ports of the MZI. Electrical signals are delivered via a custom RF probe. Voltage signals are generated with a programmable AWG, amplified, and sent to the single-pole double-throw (SPDT) RF switch. The specific electronic equipment utilized for the experiments include the Radiall R570313000 as the electromechanical SPDT RF Switch, a Liquid Instruments Moku:Pro operating as the arbitrary waveform generator (AWG), the Piezodrive PD200X operating as a multichannel, high-voltage amplifier with 20x gain, and two New Focus 1801-FC photodiodes.

We recorded both outputs of the MZI (bar and cross) for all experiments. The outputs may be normalized between 0 and 1 by simply taking the fraction in each channel and dividing it by the summed power of both channels (which also accounts for laser drift). However, in a typical measurement, fab-induced grating coupler variations and fiber array alignment may cause unequal optical coupling for the bar and cross outputs. To correct for this and obtain a more accurate measurement of the MZI state, we swept the piezoelectric phase shifter to find the maximum and minimum values for each output port. This retrieves the individual output coupling values for that iteration, which we applied to each channel before calculating the normalized power. For all measured data in the main text labeled as optical (or MZI) transmission, we applied this normalization technique.

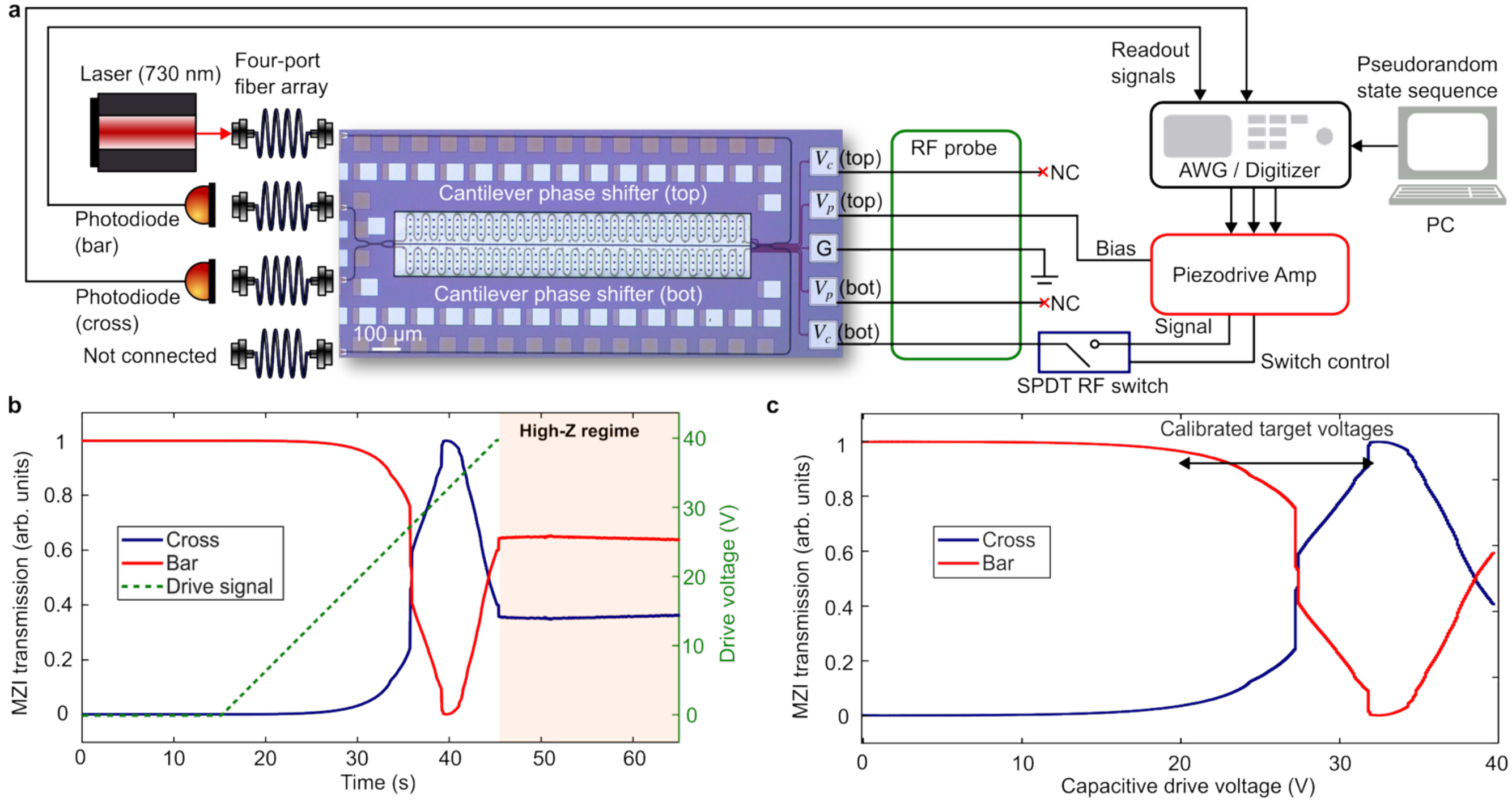

**Fig. S1: Experimental setup and memristor calibration curves for multi-bit operation.** a) Diagram of the experimental setup including optical and electronic components. b) Measured electrostatic response in time domain with ramp function terminating in high-Z. c) Derived voltage calibration curve for binary and multi-bit memristor operation.

For phase extraction, we calculated $\theta_{bar} = \pm 2\sin^{-1}(\sqrt{P_{bar}^{norm}})$, $\theta_{cross} = \pm 2\cos^{-1}(\sqrt{P_{cross}^{norm}})$. To find our initial set point, we measured the MZI response curve by using the piezoelectric phase shifter to determine if we are in the {-π,0} or {0, π} phase regime. Finally, to unwrap the phase for a smoother plot, we flipped the sign or added an offset of $2\pi$ to determine the unwrapped phase shift. This technique was primarily used to better determine the exponential relaxation of the cantilever.

**Drive voltage values for multi-bit operation:** Based on the measured DC electrostatic response curves in Fig. S1b and S1c, we used the following drive voltages in Table S1 for multi-bit operation that correspond to the desired amplitude levels which define each state.

**Table S1: Calibrated amplitude and voltage levels for 2-bit, 3-bit, and 4-bit operation**

| 2-bit calibration values | | | |
|---|---|---|---|
| State | Binary | Amplitude level | Drive (V) |
| 0 | 00 | 0.000 | 0.0 |
| 1 | 10 | 0.333 | 30.0 |
| 2 | 11 | 0.667 | 33.7 |
| 3 | 01 | 1.000 | 37.0 |

| 3-bit calibration values | | | |
|---|---|---|---|
| State | Binary | Amplitude level | Drive (V) |
| 0 | 000 | 0.000 | 0.0 |
| 1 | 001 | 0.143 | 27.3 |
| 2 | 011 | 0.286 | 29.0 |
| 3 | 010 | 0.429 | 30.0 |
| 4 | 110 | 0.571 | 31.2 |
| 5 | 111 | 0.714 | 32.4 |
| 6 | 101 | 0.857 | 33.7 |
| 7 | 100 | 1.000 | 36.2 |

| 4-bit calibration values | | | |
|---|---|---|---|
| State | Binary | Amplitude level | Drive (V) |
| 0 | 0000 | 0.000 | 0 |
| 1 | 0001 | 0.067 | 25.7 |
| 2 | 0011 | 0.133 | 27.3 |
| 3 | 0010 | 0.200 | 28.3 |
| 4 | 0110 | 0.267 | 29.3 |
| 5 | 0111 | 0.333 | 30.2 |
| 6 | 0101 | 0.400 | 30.9 |
| 7 | 0100 | 0.467 | 31.6 |
| 8 | 1100 | 0.533 | 32.2 |
| 9 | 1101 | 0.600 | 32.7 |
| 10 | 1111 | 0.667 | 33.1 |
| 11 | 1110 | 0.733 | 33.5 |
| 12 | 1010 | 0.800 | 33.8 |
| 13 | 1011 | 0.867 | 34.4 |
| 14 | 1001 | 0.933 | 34.9 |
| 15 | 1000 | 1.000 | 36.5 |

For 5-bit operation, the amplitude and corresponding voltage drive values are listed in Table S2. We note the voltages for the 5-bit experiments are slightly higher than the voltages used for the previous multi-bit experiments. The 5-bit test occurred after the >1 billion endurance test, which may have caused some degradation of electrostatic actuator.

**Table S2: Calibration voltages for 5-bit operation**

| **5-bit calibration values** | | | | | | | |
|---|---|---|---|---|---|---|---|
| State | Binary | Amplitude level | Drive (V) | State | Binary | Amplitude level | Drive (V) |
| 0 | 00000 | 0.000 | 0.0 | 16 | 11000 | 0.516 | 37.1 |
| 1 | 00001 | 0.032 | 30.2 | 17 | 11001 | 0.548 | 37.4 |
| 2 | 00011 | 0.065 | 31.3 | 18 | 11011 | 0.581 | 37.7 |
| 3 | 00010 | 0.097 | 32.1 | 19 | 11010 | 0.613 | 38.0 |
| 4 | 00110 | 0.129 | 32.6 | 20 | 11110 | 0.645 | 38.3 |
| 5 | 00111 | 0.161 | 33.1 | 21 | 11111 | 0.677 | 38.6 |
| 6 | 00101 | 0.194 | 33.5 | 22 | 11101 | 0.710 | 38.9 |
| 7 | 00100 | 0.226 | 34.0 | 23 | 11100 | 0.742 | 39.2 |
| 8 | 01100 | 0.258 | 34.4 | 24 | 10100 | 0.774 | 39.7 |
| 9 | 01101 | 0.290 | 34.9 | 25 | 10101 | 0.806 | 40.0 |
| 10 | 01111 | 0.323 | 35.3 | 26 | 10111 | 0.839 | 40.4 |
| 11 | 01110 | 0.355 | 35.7 | 27 | 10110 | 0.871 | 40.6 |
| 12 | 01010 | 0.387 | 36.0 | 28 | 10010 | 0.903 | 40.9 |
| 13 | 01011 | 0.419 | 36.2 | 29 | 10011 | 0.935 | 41.5 |
| 14 | 01001 | 0.452 | 36.4 | 30 | 10001 | 0.968 | 42.2 |
| 15 | 01000 | 0.484 | 36.7 | 31 | 10000 | 1.000 | 43.6 |

**Transmitted / Received Error Matrices:** We show the full Tx/Rx error matrices for the 4-bit and 5-bit experiments reported in the main text in Fig. S2 and Fig. S3.

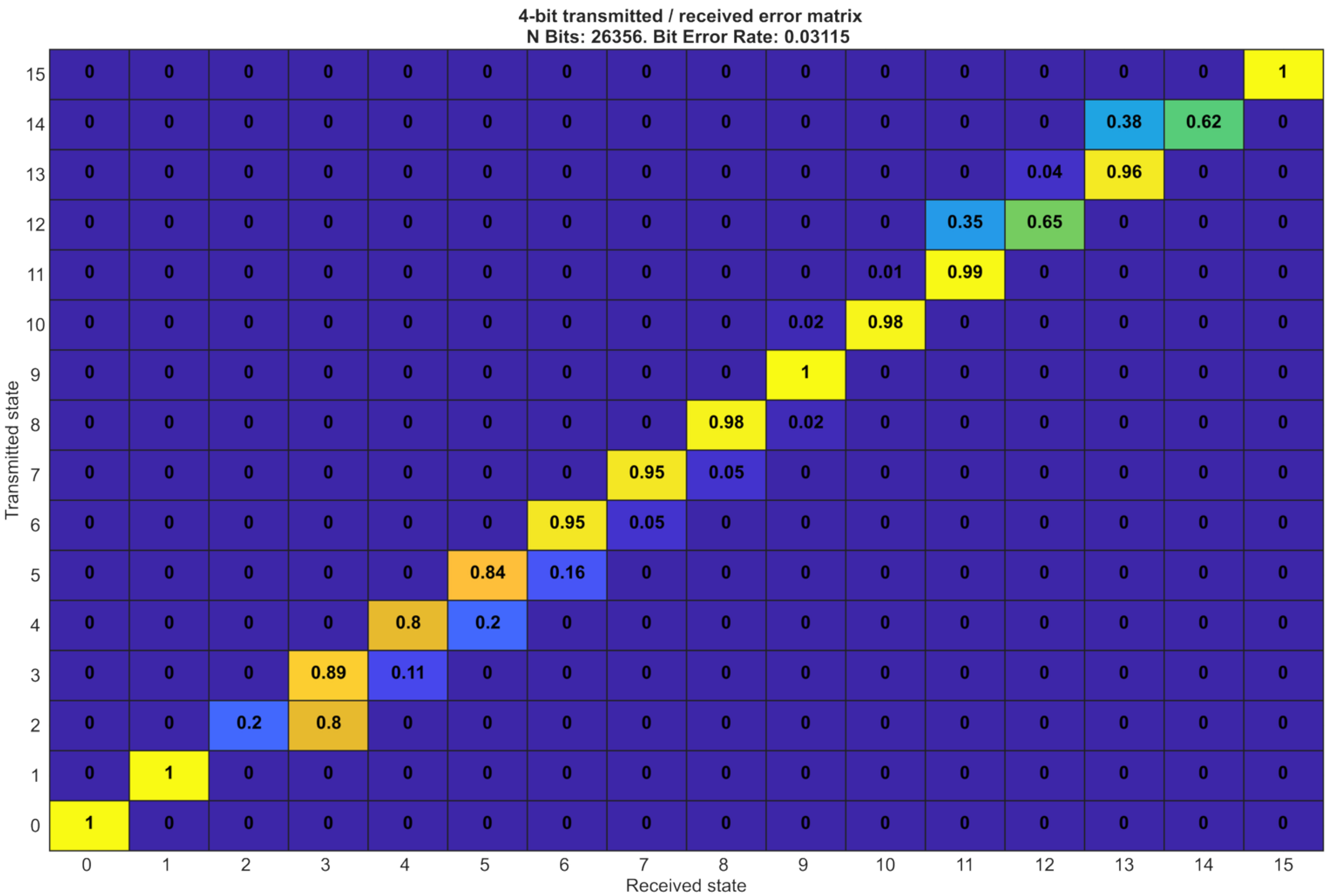

**Fig. S2: Full transmitted / received error matrix for 4-bit operation.**

Fig. S2 shows the error matrix in 4-bit operation. We observe that the error values are mostly clustered along the diagonal. All errors are contained in adjacent states (error distance of 1). Certain states such as states 12 and 14, and therefore amplitude values, are harder to achieve accurately due to the nonlinearity in the calibration curve.

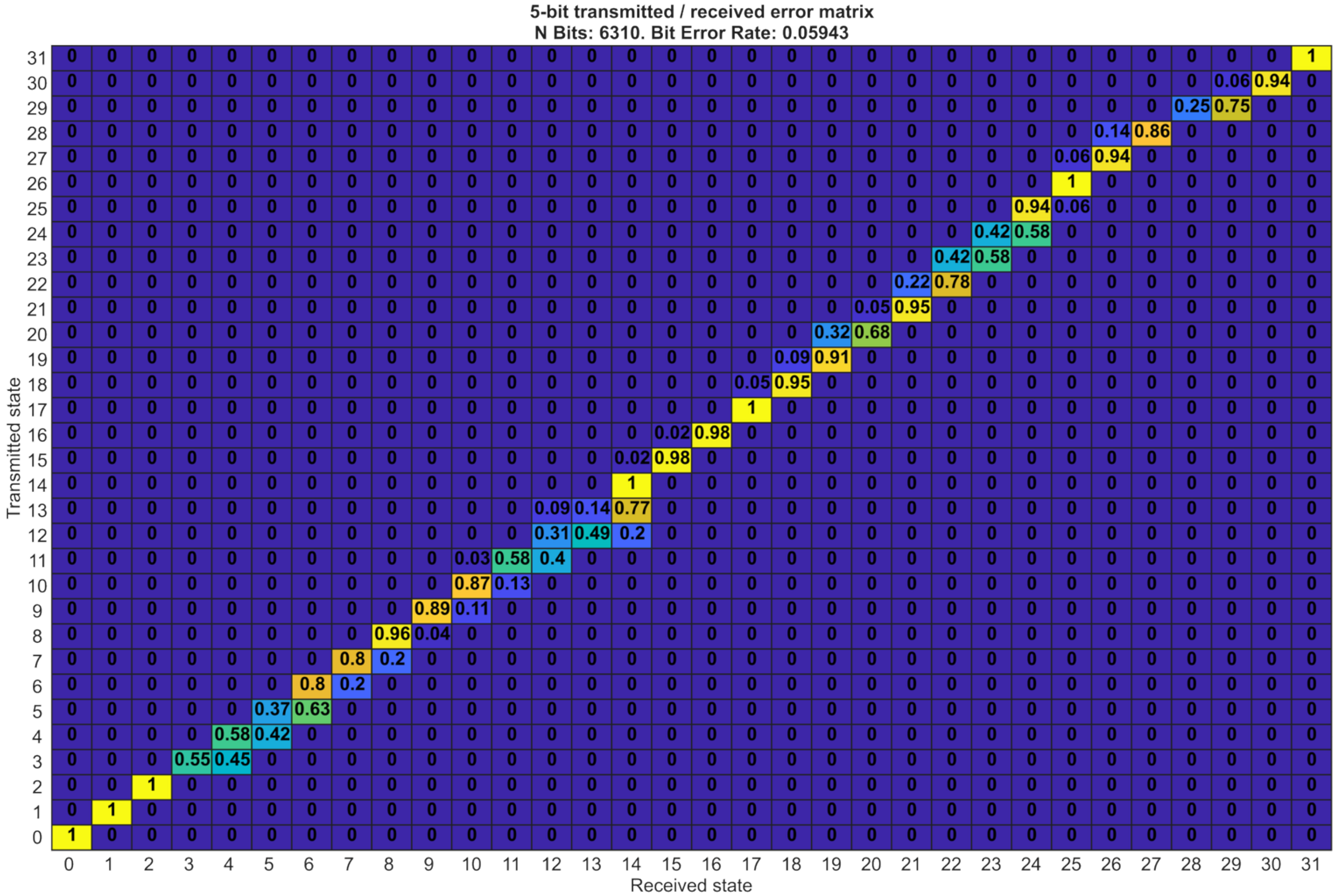

**Fig. S3: Full transmitted / received error matrix for 5-bit operation.**

The error matrix for 5-bit operation is shown in Fig. S3. Similar to the 4-bit case, some particular states such as 25-28 are particularly error prone. This is likely due to a combination of device fatigue leading to a drift between calibration and transmission. The transmission levels for these states are unstable due to the nature of the calibration curve and our encoding of all states with equal transmission ranges (excluding 0 and the highest states at the edge). This may be improved with monitoring in-the-loop that adjusts state voltage levels periodically to compensate. Likewise, we also expect major improvements with different encoding of the states that do not rely on hard-to-achieve transmission values. For example, if we retested the device and simply re-defined state 25 transmit level to that of state 26, the error rates would be 0 for the new state 25. States 26-28 may be similarly adjusted.

**Refresh circuit schematic.** Fig. S4 shows the schematic of the battery-powered electronic driver of the photonic integrated memristor. The 555 Timer is Texas Instruments TLC555CP. The electromechanical relay is S2-03PU from Sensata Technologies, Inc. The driver consists of two parts, a timing circuit and a refresh circuit. The timing circuit is designed to output a 160 ms trigger pulse every 7 seconds to the electromechanical relay. The relay is part of the broader refresh circuit which defines the memory states and associated voltages required to achieve them. Depending on the desired calibration, we used both 45 V and 54 V configurations (5 x 9 V batteries or 6 x 9 V batteries). We note that Fig. S4b shows the more general configuration of state 0 possibly assigned to an arbitrary, non-zero value. However, for all battery-powered experiments reported in the paper, we set state 0 to 0 V.

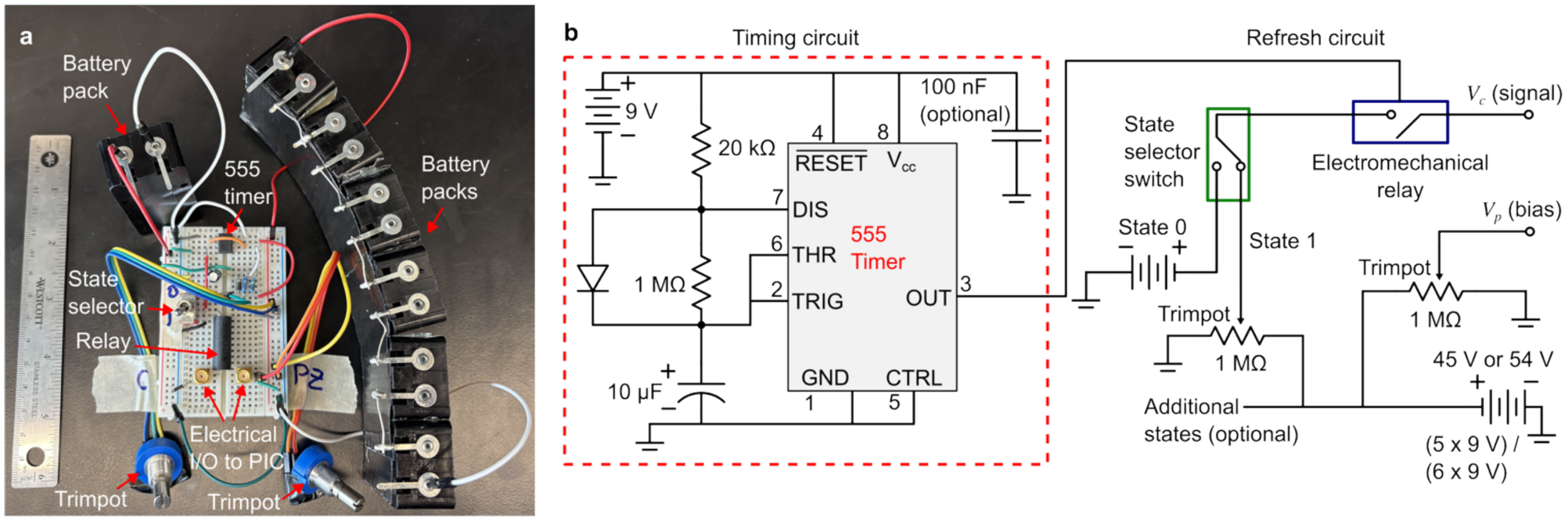

**Fig. S4: Battery-powered driver circuit for the photonic integrated memristor.** a) Photo image of the driver with a ruler for size reference. Major electrical components are labeled b) circuit diagram of the driver electronics consisting of a timing circuit and a refresh circuit.

## S2. Additional device characterization data

**Resistance and capacitance measurements:** We measured the electrical impedance of the memristor, calculated from scattering coefficients, to determine the device capacitance. We first calibrated out the vector network analyzer (Keysight E5061B) and test cables with an Electronic Calibration Module (Keysight N5772 A). Due to how the PIC is mounted, we were unable to fully calibrate out a 12" cable and PCB on our packaged device in our test setup. To alleviate this, we measured the "open" S11 response of an equivalent cable and PCB (without the PIC) and observed a residual capacitance of ~34 pF and inductance of 39 nH. We then modeled the parasitic inductance and capacitance in series with a pure capacitor for electrostatic actuators (Fig. S5a) or a modified Butterworth-Van-Dyke (BVD) model for piezoelectric actuators (Fig. S5b) for our device under test (DUT). Fitting our DUT model to measured S11 data, we found good agreement between the circuit model and our photonic actuators (Fig. S5c-e). The electrostatic capacitance was $C \sim 9$ pF on both the top and bottom arms of the electrostatic device. We also found that applying a strong electrostatic DC bias slightly increased the device capacitance to ~12 pF at 40 V as expected due to the decreasing gap between the electrostatic electrodes at higher biases. Similarly, the piezoelectric capacitance was $C_0 \sim 26$ pF with other mechanical parameters fitted to be $L_1 = 151$ µH, $C_1 = 0.36$ pF, and $R_1 = 1002$ Ω.

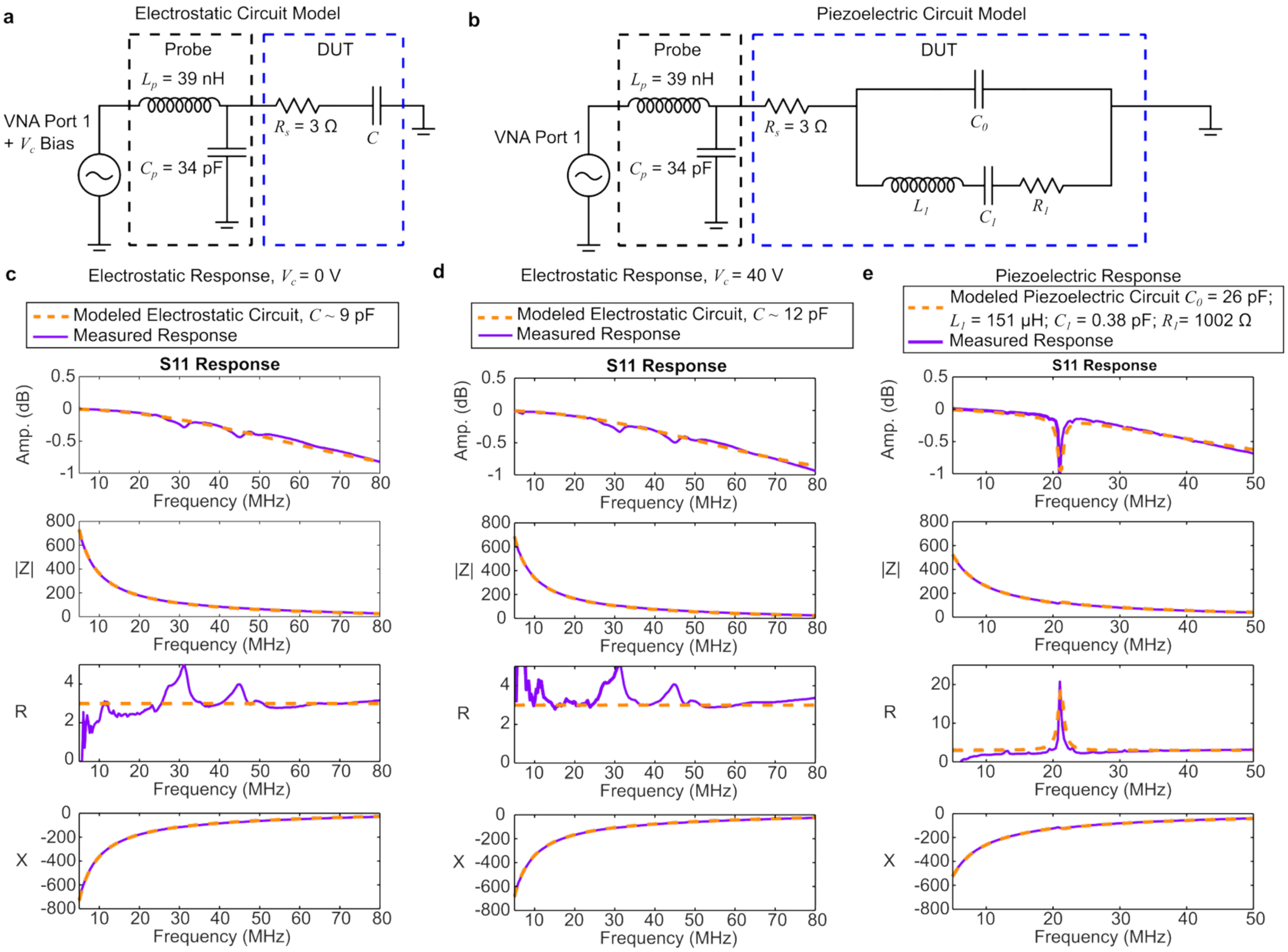

**Fig. S5: Impedance analysis of electrostatic and piezoelectric modulation.** Circuit model for the a) electrostatic capacitor and b) piezoelectric actuator. c) Electrostatic S11 response with $V_c = 0$ V with circuit model of $C \sim 9$ pF overlaid. d) Electrostatic S11 response with $V_c = 40$ V with circuit model of $C \sim 12$ pF overlaid. e) Piezoelectric S11 response, with modified BVD circuit model overlaid. $C_0 = 26$ pF, $L_1 = 151$ µH, $C_1 = 0.36$ pF, and $R_1 = 1002$ Ω.

As a reference, we estimate the capacitance with the parallel plate capacitor equation $C = \frac{\epsilon A}{d}$ where $A$ = 1232 μm x 91 μm, $d$ = 800 nm, where roughly 600 nm is oxide and 200 nm is air. Using a weighted average and $\epsilon_{ox}$ = 3.9$\epsilon_0$, we obtain $\epsilon \sim 3.2\epsilon_0$ and $C \approx$ 4 pF. The discrepancy likely due to small errors in the measurement, imperfect calibration of the circuit, and fabrication variations in the exact layer thicknesses.

We also performed an IV measurement shown in Fig. S6 for both the electrostatic and piezoelectric voltage inputs. Fig. S6a shows the electrostatic IV starts flat but becomes fairly linear past ~25 V, fitting to a leakage resistance of ~215 GΩ. Conversely, the piezoelectric arm (Fig. S6b) had much more leakage current and a nonlinear, diode-like IV curve. The resistances from 25 V onwards varied from ~3 GΩ to ~66 GΩ. We estimate the hold power of the single-arm piezoelectric actuator ($V_\pi$ ~ 36 V) to be 40 nW at $\pi$ phase shift, drawing about 1.1 nA. In push-pull configuration, the hold power would be greatly reduced to 3.6 nW at $\pi$ phase shift, where each phase shifter holds 18 V and draws 0.1 nA. We estimate hold power of the single-arm electrostatic actuator at the maximum 35 V for memristor operation is 5.7 nW. The energy required to reconfigure the capacitor is approximately (assuming no recycling of stored capacitor energy) $CV^2 = 12$ nJ.

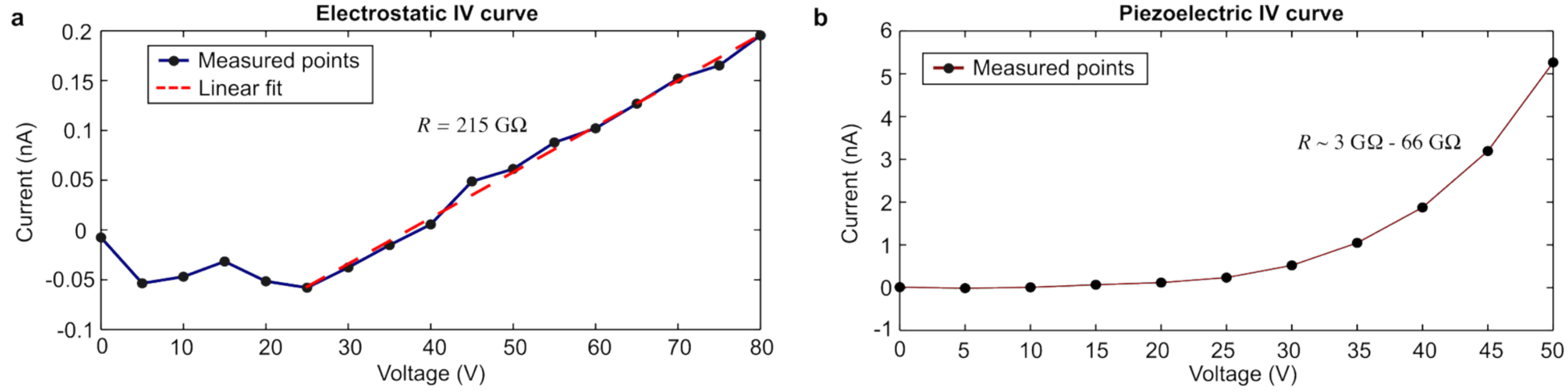

**Fig. S6: IV curve measurements of our primary test memristor.** a) IV curve for the electrostatic capacitor, showing a linear region >25 V with a fit of R ~ 215 GΩ. b) IV curve for the piezoelectric actuator, showing decreasing resistance from 66 GΩ at low voltage to ~3 GΩ at higher voltages.

**Long-term phase storage and decay.** We present additional data to on the nature of the phase storage and decay. We measured the phase decay on a different device of the same design as that of the main text. Fig. S7a plots the normalized transmission of the long-term decay over several days with the device in high-Z. The initial voltage ramp and MZI phase response are shown more clearly in in Fig. S7b, where the phase undergoes an increase then decrease throughout the ramp. Fig. S7c shows the phase decay over the entire measurement. Finally, assuming a 1-to-1 voltage-to-phase relation as the device decays, we modeled an exponential voltage-and-phase decay and overlaid the resulting decay curve to the measured data, shown in Fig. S7d. We observe that, for an exponential lifetime of $\tau$ = 110000 s, the expected phase decay matches well of that measured experimentally. However, after about 20 hours the measured phase decays faster than the simple decay model predicts.

The data here suggests a few interesting things. The first is the decay lifetime $\tau$ far exceeds the $RC$ constant based on resistance and capacitance measurements above by 3 to 5 orders of magnitude. The second is the decay is not strictly exponential, but has abrupt jumps as it goes through the decay path. This strongly indicates the storage mechanism has a substantial mechanical component, not purely electrical, likely due to a combination of mechanical potential energy and van der Waals forces [21,29]. We will investigate the storage physics in more detail in a future study.

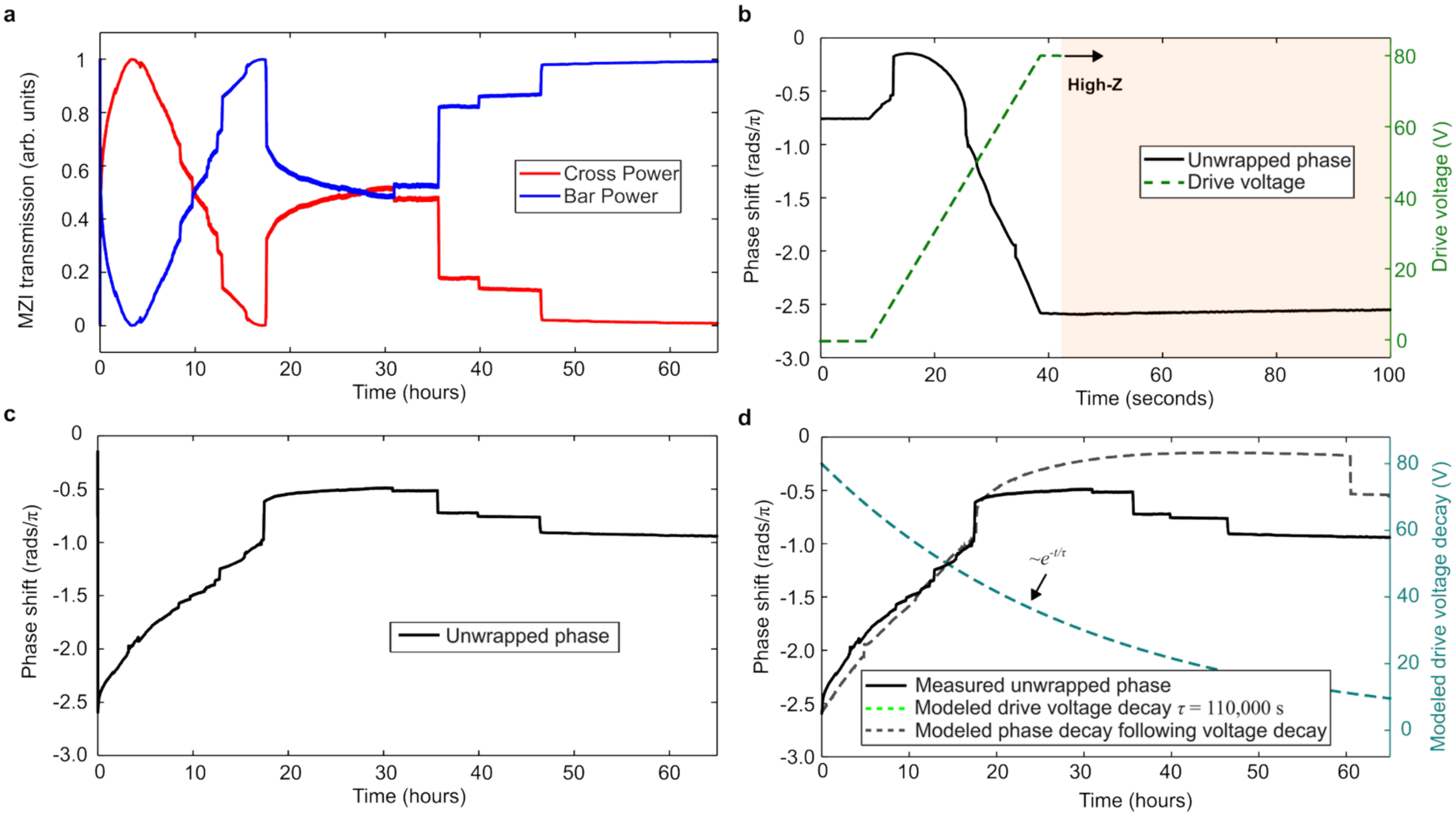

**Fig. S7: Long phase decay measurement in high-Z.** a) Normalized transmission through the MZI showing slow decay over several days b) Zoomed-in plot of the drive voltage and wrapped phase response in the first 100 s of the long decay experiment. c) Unwrapped phase decay data plotted over the same duration. d) Unwrapped phase decay overlaid with an expected phase decay response assuming an exponential decay model and a one-to-one voltage-phase relation reversed to the original state.

**Bit resolution vs refresh time.** We note that while for binary operation the refresh rates may exceed 1000 s, multi-bit operation will suffer bit errors before this value is reached. Fig. S8 shows an example 3-bit operation at 100 s refresh rate (10x longer than the reported main text data) that starts to exhibit bit error rates of 0.078. Higher bit operation will require shorter refresh times. However, it is possible to improve this by defining a more robust operating point, i.e. re-calibrating the set voltages to not the middle of the state amplitude range but on the higher end to account for the decay.

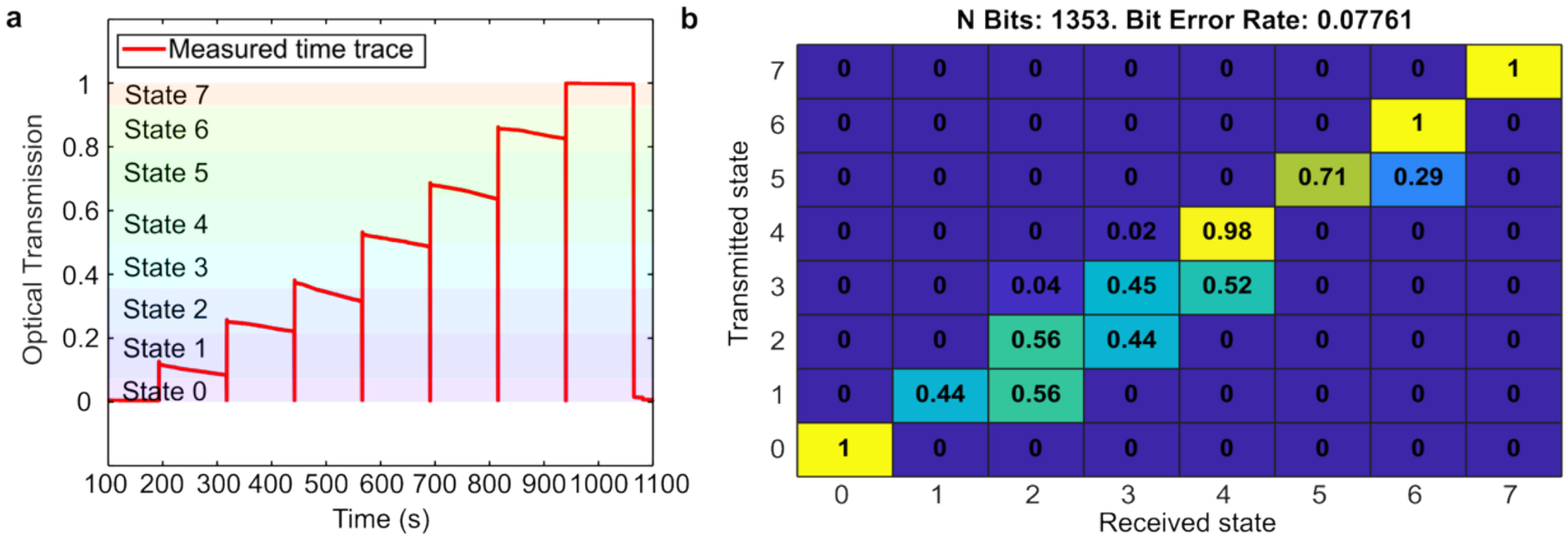

**Fig. S8: 3-bit (8 state) operation at 100 s refresh rate.** a) Measured time trace of a linear step sequence showing some states decaying out of the state amplitude range at 100 s write-read repetition period. b) Transmitted / received error matrix, showing a BER 0.078, compared to 0 BER at 10 s write-read repetition period.

**Alternate cantilever geometries.** We also investigated a different geometry cantilever with a much longer overhang of 220 µm with the same electrostatic-piezoelectric actuation layers, shown in Fig. S9. These devices did not have as strong of an electrostatic response, but still exhibited the long-lived memory observed in the main reported device. The data confirms the stiction behavior is dependent upon the geometry. The exact optimal geometry for maximizing the electrostatic memory and actuation will be the subject of a future investigation.

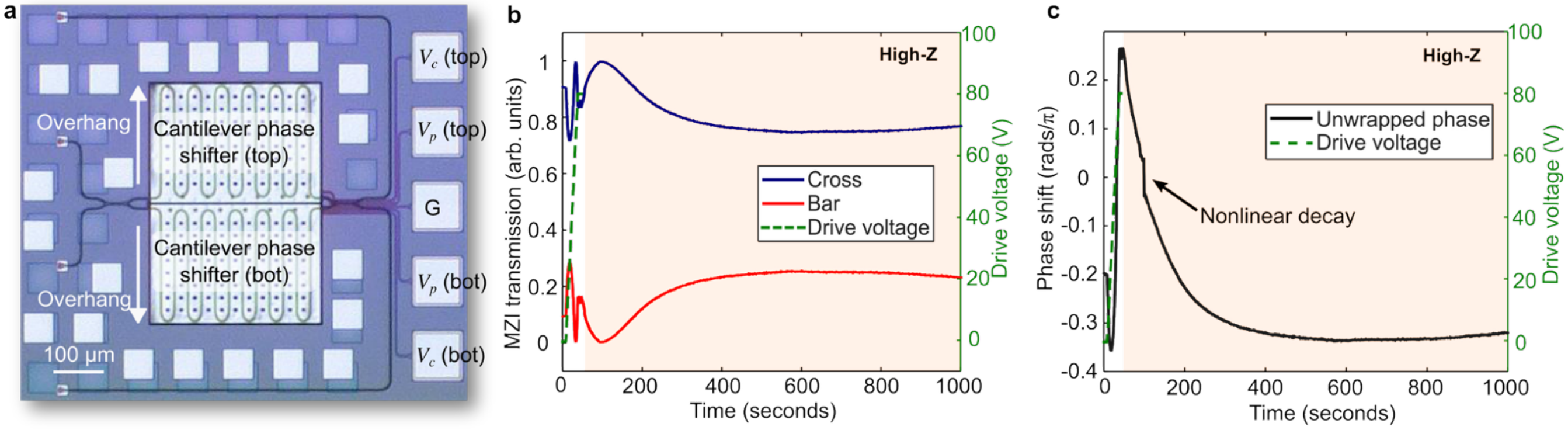

**Fig. S9: Characterization of non-volatility in a 220 µm overhang cantilever phase shifter.** a) Microscope image of the cantilever memristor geometry with cantilevers and electrical I/O labeled. b) Measured optical transmission of the device up to 80 V with slow decay once switched to high-Z. c) Unwrapped phase decay over time.

### S3. Energy efficiency analysis

We explore the energy expenditures of using our integrated memristor to implement programmable weights as part of a simple neural network under various use-cases. As a baseline, we compare our device to the ubiquitous thermo-optic implementations as well as electrically programmed phase change materials (PCMs). Other systems such as ferroelectric BTO or optically programmed PCMs were considered but we excluded them as it was less straightforward to estimate the system energy use for a single reprogram due to necessary procedures such as re-poling and generation of short-pulse lasers.

From our energy-use values calculated in section S1, we define the programming energy to be approximately 12 nJ and a hold-energy of 5.7 nW. We immediately see that a refresh rate of greater than ~2 seconds will result in energy savings, with further proportional decreases as refresh rate decreases.

The average power given a reprogram rate *T* is calculated as

$$P_{avg} \approx \frac{E_p}{T} + P_{hold}$$

where $P_{hold}$ is the hold power for a particular modulator technology (0 for memristors), $E_p$ is the total energy required to reprogram the modulator or memristor. Table S3 shows the average power consumption of electrically programmed technologies under various reprogram rates.

**Table S3: Average power consumption of memristor use**

| Technology | Ref. | $E_p$ | $P_{avg}$ ($T = 2\ s$) | $P_{avg}$ ($T = 10\ s$) | $P_{avg}$ ($T = 100\ s$) | $P_{avg}$ ($T = 1000\ s$) | $P_{avg}$ ($T =$ 24 hrs) |
|---|---|---|---|---|---|---|---|
| Thermo-optic | [34] | - | 3 mW | 3 mW | 3 mW | 3 mW | 3 mW |
| PCM (graphene) | [7] | ~1 µJ* | 500 nW | 100 nW | 10 nW | 1 nW | 10 pW |
| PCM (Si) | [8] | ~4 µJ* | 2000 nW | 400 nW | 40 nW | 4 nW | 40 pW |
| MEMS always on | This work | 12 nJ | 12 nW | 6.9 nW | 5.8 nW | 5.7 nW | 5.7 nW |
| MEMS with non-volatility | This work | 12 nJ | 6.0 nW | 1.2 nW | 120 pW | 12 pW | 12 pW† |

*includes both SET and RESET pulse energies

†assumes a 1000 s refresh rate

We observe that, at 1000 s refresh rate, our MEMS uses energy at a rate close to a PCM memristor reprogrammed approximately once per day. In other words, for practical use (executing matrix-vector multiplications, for example), PCMs are still more energy efficient if the device is not programmed or recalibrated more than once per day. Otherwise, for more frequent programming rates, MEMS with refresh would be more energy efficient. We add that the device utilization such as inference must be occurring for the entire duration of the 24 hours for this comparison to be valid (100% hardware uptime); otherwise the MEMS with refresh devices can be shut off to further save energy.